\documentstyle[11pt,jkas2]{article}

\newcommand{\AmS}{{\protect\the\textfont2
  A\kern-.1667em\lower.5ex\hbox{M}\kern-.125emS}}

\title{RECENT PROGRESS IN STRING INFLATIONARY COSMOLOGY}
\author{SOO-JONG REY}
\address
{Physics Department, Seoul National University, Seoul 151-742 KOREA\\
Department of Physics, Stanford University, Stanford CA 94305 USA\\
Stanford Linear Accelerator Center, Stanford University, Stanford CA 94309
USA}
\begin{document}
\setlength{\arraycolsep}{0pt}
\begin{titlepage}
\begin{flushright} SNUTP 96-087 \\ {\tt hep-th/9609115}
\end{flushright}
\vfill
\begin{center}
{\large\bf  RECENT PROGRESS IN STRING INFLATIONARY COSMOLOGY${}^\dagger$}\\
\vskip 7.mm
{\large \bf SOO-JONG REY}\\
\vskip 5.mm
{\sl Physics Department, Seoul National University, Seoul 151-742 KOREA} \\
{\sl Department of Physics, Stanford University, Stanford CA 94305 USA}\\
{\sl Stanford Linear Accelerator Center, Stanford University, Stanford
CA 94309 USA}
\vfill
{\bf ABSTRACT}
\end{center}
\begin{quote}
Super--inflation driven by dilaton/moduli kinetic energy is naturally
realized in compactified string theory. Discussed are selected topics of
recent development in string inflationary cosmology: kinematics of 
super-inflation, graceful exit triggered by quantum back reaction, and 
classical and quantum power spectra of density and metric perturbations.

\vfill      \hrule width 5.cm
\vskip 2.mm
{\small\small
\noindent $^\dagger$ 
Invited Talk given at 7th Asian-Pacific Regional Meeting of the International
Astronomical Union, August 19 - 23, 1996 Pusan, Korea. To be published in the
Proceedings.  }  
\end{quote}
\end{titlepage}

\abstract{
Super--inflation driven by dilaton/moduli kinetic energy is naturally
realized in compactified string theory. Discussed are selected topics of recent
development in string inflationary cosmology: kinematics of super-inflation,
graceful exit triggered by quantum back-reaction, and classical and quantum
power spectra of density and metric perturbations.
}
\maketitle
\setcounter{footnote}{0}

%
  


\author{SOO-JONG REY} 
\address{Physics Department, Seoul National University, Seoul 151-742 KOREA}
\address{Department of Physics, Stanford University, Stanford CA 94305 USA}
\address{Stanford Linear Accelerator Center, Stanford University, Stanford 
CA 94309 USA}

\abstract{
Super--inflation driven by dilaton/moduli kinetic energy is naturally
realized in compactified string theory. Discussed are selected topics of recent
development in string inflationary cosmology: kinematics of super-inflation,
graceful exit triggered by quantum back-reaction, and classical and quantum
power spectra of density and metric perturbations.
}


\maketitle  

\section{MOTIVATIONS FOR STRING INFLATION}           
Despite impressive success, the Standard Model of big--bang cosmology
is known to suffer from two `naturalness' problems: observed homogeneity and 
spatial flatness of the present universe that cannot be explained in a 
natural way. 
Rather they point to a period of inflationary expansion in the past.  
The evolution of Friedmann-Robertson-Walker (FRW) scale factor is governed
by the Einstein equations
\begin{eqnarray}
\Big({\dot a \over a} \Big)^2 &=& {8 \pi \over 3} G \rho - {k \over a^2}
\label{hubble} \\
{\ddot a \over a} &=& -{4 \pi \over 3} G (\rho + 3p)
\label{graveqn}
\end{eqnarray}
where $\rho$ and $p$ are the energy density and the pressure of the matter 
and $k$ measures the spatial curvature.
The spatial flatness problem is solved naturally if the matter density has 
grown much larger than the spatial curvature.
Thus inflation is characterized by a period
during which the ratio ${8 \pi \over 3} G \rho / (1/a^2) = {\dot a}^2 + k$ 
has increased with time, viz., $\dot a > 0$ and $\ddot a > 0$.
Eq.(\ref{graveqn}) shows that such an accelerated expansion is 
possible only for exotic matter satisfying 
$ \rho + 3p < 0$.
Inflation solves the horizon problem automatically. The physical distance 
for a fixed comoving separation scales as $a$. 
The cosmic horizon is inversely proportional to the Hubble parameter. 
Thus, during inflation, their ratio $a / (a / {\dot a}) = \dot a$ 
grows with time since $\ddot a > 0$. 
This implies that the physical distance scale is stretched outside
the horizon so that the correlation encompasses an enormous spatial volume,
hence, solves the horizon problem. 

There are three possible types of inflation. The first,
de Sitter inflation $a(t) = \exp (H t)$, arises from nonzero vacuum energy 
during weakly first-order phase transition. 
The second, power-law inflation $a(t) = t^p$ for $t> 0, p >1$ arises in
many models of supergravity with exponential potential. The third,
super-inflation $a(t) = (-t)^p$, $t> 0, p < 0$ is the least familiar one
but arises for Brans-Dicke-type gravity theories including string theory.
The novelty of the super-inflation is that it is driven by kinetic energy 
rather than vacuum potential energy as is required for the former two 
inflations. This is gratifying since it
has been known that the vacuum potential energy has to be fine-tuned
in order to achieve an observationally successful inflation. 
As such, for inflations driven by potential energy, the naturalness problems
of observational cosmology have been traded for the naturalness problems of
underlying microscopic physics.

\section{KINEMATICS OF STRING SUPER-INFLATION} 
Classical string dynamics at sub-Planck scale is described by a Brans-Dicke
type effective Lagrangian expressed as a power--series expansion of spacetime
derivatives:
\begin{eqnarray}
L &=& e^{-2\phi} [- R - 4 (\nabla \phi)^2 + (\nabla T)^2 + \cdots] 
\nonumber \\
&+& L_{\rm matter}.
\end{eqnarray}
The ellipses denote (classical) higher-derivative terms,
$L_{\rm eff}$
is the Lagrangian describing matter coupling, and
$\phi$ and $T$ are dilaton and moduli fields (associated with
the size and the shape of compactified space) respectively. 
Eq.(3) indicates that, 
in string theory, the Newton's constant is not a fixed quantity but 
determined dynamically by the dilaton $\phi$: $G = e^{2 \phi} / 16 \pi$.

As first pointed out by Veneziano (Veneziano, 1991), the 
string theory gives rise to the super-inflation naturally.
Veneziano has shown that Einstein equation and $\phi, T$ equations of 
motion derived from Eq.(3) have always two cosmological branches. 
For example, for vacuum without matter, the first branch 
exhibits decelerating expansion and growing Newton's constant
\begin{equation}
a(t) = t^{+ 1/\sqrt 3}, \hskip0.3cm e^\phi = t^{-1 + \sqrt 3}, 
\hskip0.5cm t > 0,
\end{equation}
while the second branch represents accelerating expansion and growing
Newton's constant
\begin{equation} 
a(t) = (-t)^{-1/\sqrt 3}, \hskip0.3cm e^\phi = (-t)^{-1-\sqrt 3}, 
\hskip0.5cm t < 0.
\end{equation}
Similarly, for $p = \pm \rho/3$ matter,  
the first branch  ($p = +\rho/3$) represents a radiation-dominated FRW 
universe and  frozen Newton's constant
\begin{equation}
a(t) = t^{1/2}, \hskip0.3cm \phi = {\rm constant}, \hskip0.5cm t > 0.
\end{equation}
The second branch ($p = - \rho/3$) represents the universe with accelerated 
expansion and growing Netwon's constant
\begin{equation}
a(t) = (-t)^{-1/2}, \hskip0.3cm \phi = -3 \log (-t), \hskip0.5cm t < 0.
\end{equation}

Veneziano has shown that the two branches are related 
each other by simultaneous time-reversal $t \rightarrow -t$ and 
`scale-factor duality': 
$a \rightarrow 1/a, \,\, \phi \rightarrow \phi - 6 \log a$ and 
$(p/\rho) \rightarrow - (p/\rho)$.
The duality is a consequence of the underlying string theory 
symmetries, hence,
is the most distinguishing feature of string cosmology from the others.
More interestingly the scale-factor duality may offer a stringy mechanism 
for exiting from inflation: by duality
the branch can flip from the super-inflation to the FRW-type one.  
For physically sensible branch change, metric, Newton's constant and all
other physical quantities should interpolate smoothly across the moment of 
fixed point of the scale-factor duality $t^*$: $a(t^*) = 1/a(-t^*), \,\, 
\phi (-t^*) = \phi(t^*) - 6 \log a(t^*)$. 

Kinematical distinction of the super-inflation compared to the other 
two inflations is most clearly seen from the behavior of the cosmic horizon
(Fabbri et.al., 1985).
The particle horizon given by the inverse Hubble parameter
$R_H = 1/ H = (-t)/p$ shrinks to zero size asymptotically as 
$t \rightarrow 0^-$.
In de Sitter inflation $R_H = 1/H$ remains frozen, while
in power-law inflation $R_H = t/p$ grows large. This distinction bears
an important consequence to the generation of the primordial density and
metric perturbations. 
The physics behind generating these perturbations is the quantum 
fluctuation produced inside the subluminal horizon. Once produced, the quantum
fluctuation is streched outside the horizon and behaves as classical
density and metric perturbations (Rey, 1987).
Since the horizon for super-inflation shrinks during inflation, 
quantum fluctuations are parametrically squeezed
toward shorter wavelengths. More precisely, the first horizon-crossing
condition implies that the higher frequency modes cross the horizon at later
time. With increasing Hubble parameter in time, the power spectra of 
the higher frequency modes are amplified with respect to lower frequency mode 
ones. This has an important bearing on the formation of the large-scale 
structures as discussed later.

\section{GRACEFUL EXIT VIA QUANTUM BACK REACTION}
As mentioned the string theory offers an extremely attractive 
possibility of exit from inflation utilizing the scale-factor duality. 
Unfortunately, extensive study
has shown it impossible (Brustein and Veneziano, 1994) because the curvature 
and the Newton's constant have turned out discontinuous and
divergent at the fixed point moment $t^* = 0$.
This problem seems very generic and is now known as the `graceful exit problem'
of string inflationary cosmology.

On a closer look, however, the problem arises entirely within the
classical approximation. Near the fixed point moment, both the curvature and 
the Newton's constant grow indefinitely so that higher-order curvature 
and quantum corrections become important. Therefore, whether a smooth 
transition between the two branches is possible or not should be determined 
only after a full-fledged quantum stringy effect is taken into account.

Antoniadis, Rizos and Tamvakis (Antoniadis et.al., 1994) 
have initiated the study of the quantum effect to the
graceful exit problem. They have found that under certain initial conditions 
the quantum back reaction of the fluctuating matter leads to a smooth 
transition from the super-inflation branch to the FRW-type branch.
Their analysis has been further extended (Easther and Maeda, 1996) 
for nonzero spatial curvature and have reached essentially the same 
conclusion. 

In fact, it is
possible to obtain an `exact' quantum analysis for a two-dimensional 
truncation (Rey, 1996), which still captures 
all the underlying essential physics of four dimensions.
After the truncation, the classical Lagrangian is given by
\begin{equation}
L_{classical} = e^{-2 \phi} (-R - 4 (\nabla \phi)^2)
+ {1 \over 2} (\nabla {\vec f})^2 
\label{2daction}
\end{equation}
where $\phi$ and ${\vec f}$ refer to the dilaton and the $N$-component 
(Ramond-Ramond) matter field. 
By solving the equations of motion, two branches are found exactly. The 
super-inflation branch:
\begin{equation} 
(ds)^2 = [d \tau^2 - ({{\tilde M} \over - \tau})^2 dx^2]; 
\hskip0.3cm -\infty < \tau \le 0
\end{equation}
with a growing dilaton: $\phi = - \log (-2\tau)$, and
the FRW-type branch:
\begin{equation} 
(ds)^2 = [ d\tau^2 - (M \tau)^2 dx^2]; 
\hskip0.3cm 0 \le \tau < \infty 
\end{equation}
with a frozen dilaton.

The two branches also show discontinuous and divergent 
curvature and Newton's constant at the fixed point moment $\tau^* = 0$, 
hence, a two-dimensional version of the graceful exit problem. 
In two dimensions, the quantum corrections are entirely specified by 
the conformal anomaly arising at one loop only, hence, exactly solvable 
for a given spin and multiplicity content of the massless fields. It is 
found that 
\begin{equation}
L_{\rm quantum} = L_{\rm classical} 
+ {\kappa \over 2} \big[ R {1 \over \partial^2 } R 
+ 2 \phi R \big]
\label{effaction}
\end{equation}
where $\kappa = (N-24)/24$ and is assumed negative definite (this last 
condition has been relaxed for a more general class of truncations 
(Gasperini and
Veneziano, 1996)). For detailed analysis of the quantum effects, 
we refer to the original work (Rey, 1996). Here, we sketch 
the main result. Due to the quantum corrections, both
the scale factor and the dilaton evolves beyond the classically allowed
 region and interpolates between the two branches.
A straightforward calculation shows the scalar curvature evolves as 
\begin{equation}
R= 16 e^{2 \phi} / (1 + |\kappa| e^{2 \phi}/2)^3,
\end{equation}
which vanishes at past and future infinity at which $\phi \rightarrow \pm 
\infty$. Hence, the classical singularity at $\tau = 0$ is now completely 
erased out and the inflation has ended gracefully! The $\kappa$ dependence
of the curvature clearly indicates that the graceful exit is a quantum 
mechanical effect.

Similar conclusion is reached for the quantum corrected string
vacua in four dimensions. 
The scalar curvature again vanishes at asymptotic past/future 
infinity but approaches a finite positive maximum at the fixed pont moment
$\tau = 0$. We thus conclude that quantum corrected string theory 
resolves the classical singularity and exit super-inflation 
gracefully.

\section{POWER SPECTRA OF DENSITY AND GRAVITATIONAL WAVES}
Further indication that the quantum back reaction is an essential element 
for a successful string inflation 
comes from the constraints of the large-scale 
observational cosmology. 

As emphasized above, if the quantum back reaction effect is ignored, 
the cosmic horizon shrinks with time during the super-inflation 
epoch. The shrinking horizon amplifies quantum fluctuations parametrically
as they are stretched outside the horizon. Because of this effect, it is
expected that the power spectra of primordial scalar and tensor perturbations 
are enhanced characteristically to higher frequency than the spectra for 
de Sitter or power-law inflations. Explicit calculations 
(Brustein et al., 1995, Hwang, 1996) have confirmed this expectation. 
The power spectrum at the moment of re-entrance inside the horizon
$aH|_{\rm HC} = k$ during matter-dominated epoch is given by
\begin{equation}
P(k, t_{\rm HC}) = {(a H)^4 \over 2 \pi^2} {|\delta (k)|^2 \over k}.
\end{equation}
Here $|\delta (k, t)|^2 \equiv A(t) k^n$ denotes the conventionally normalized
power spectrum of the density contrast $\delta \equiv \delta \rho / \rho_o$. 
Up to logarithmic corrections, the spectral index is found to $n=4$, hence, 
strongly tilted to higher frequency modes. This should be contrasted to the 
observationally supported near-Harrison-Zeldovich spectrum $n \approx 1$. 
Density perturbation with such a high spectral index is problematic to seed
the large-scale structure formation. For instance, consider the temperature
fluctuation of the cosmic microwave background radiation
$\delta T(\Omega_2) / T = \sum_{l,m} a_{lm} Y_{lm}(\Omega_2)$
induced by the scattering off the gravitationl potential perturbations.
Power spectrum of the $l$-th spherical mode is given by 
\begin{equation}
|a_l|^2 = \pi \int_0^\infty {dk \over k} \Big(j_l ({2k \over H_0})\Big)^2
P(k).
\end{equation}
For the spectral index $n=4$ the short wavelength contribution is
so large that Eq.(14) diverges for all $l$. Recent COBE observations clearly 
contradicts this, hence, seems to rule out the string super-inflation as a seed
for the large-scale structure formation. 

However, the above calculations have not taken into accout of 
the important quantum mechanical
effects to the dynamics of the horizon during inflation.
Especially, since the relevant density perturbations at the present 
large-scale observations have left the horizon near the very end of the 
super-inflation epoch, the quantum back reactions should have
become significant by then. Eq.(12) shows that the back reaction tends to 
retard the rate the horizon shrinks.  It is now easy to understand that any 
change of horizon dynamics affects directly the shape of the power spectra.

The super-inflation has started essentially classically initially. Therefore 
low-frequency fluctuations generated during earlier stage should show 
the characteristic $n=4$ spectral index.  
On the other hand, toward the end of inflation, the 
back reaction has slowed down significantly the rate the horizon shrinks 
to the point $\dot H \rightarrow 0$, hence, evolves de Sitter-like essentially.
Therefore the spectral index of higher frequency quantum fluctuations that 
have left the horizon at this later stage should be close to that of the 
de Sitter inflation, viz., $n \approx 1$.
The crossover from the classical, low-frequency regime to the quantum 
mechanical, high frequency regime takes place at some intermediate scale 
$k = k^*$, and is model-dependent.
As a result, the fully quantum corrected power spectra of density perturbation
should exhibit frequency-dependent spectral index $n(k \ll k^*) \approx 4$, 
$n(k \gg k*) \approx 1$ which 
interpolates monotonically between the two limits.
Consequently the quantum effects keep CMBR partial wave
power spectra Eq.(14) from diverging at high frequency.

The gravity wave power spectrum can be calculated in a similar
manner and exhibit classically a strongly tilted spectra with $n=4$.
Again quantum back reaction will curb down the spectral index 
at higher frequency regime, hence, the actual gravitational wave
signal would not be as strong as what the classical power spectra shows.

The above discussions clearly point to the importance of calculating 
fully quantum corrected power spectra for density and metric 
perturbations. In addition,
stochastic dynamics (Rey, 1987) of inflaton during the super--inflation
exhibits distinct non-Gaussian signals from the de Sitter or power-law 
inflations. A full exposition of these calculations will be reported 
elsewhere.

In this talk I have summarized basic features of the string inflationary 
cosmology. I have emphasized that an essential element to string cosmology
is the full-fledged quantum back reaction effect.
The features should confront present and future observations 
and experiments. The prospect is quite exciting for
both string theorists and observational cosmologists. 
For string theorists, observational cosmology offers the first direct
observation of relic cosmological string effects .
For observational cosmologists, string theory offers the first natural
model of inflationary cosmology and unique signature of relic density
and gravitational waves.

\acknowledgements
I acknowledge discussions with M. Gasperini, J.-C. Hwang and G. Veneziano.
This work was supported in part by NSF-KOSEF Bilateral Grant, KOSEF
Purpose-Oriented Grant 94-1400-04-01-3 and SRC Program, KRF International
Collaboration Grant and Non-Directed 
Research Grant, Ministry of Education BSRI 95-2418,
and Seoam Foundation Fellowship.

\end{document}